# Does ESG Consistently Promote the Corporate Financial Performance? A Study of the Global Cruise Industry

**Yuechen Wu**[1,a,*]
[1] *College of Transport and Communications, Shanghai Maritime University, Shanghai, China*
*a. 202110220056@stu.shmtu.edu.cn*
**corresponding author*

**Abstract:** The analysis of determinants of a company's financial performance has aroused significant attention, particularly, the environmental, social, and governance (ESG) has been the research focus in recent years. In addition to increasing revenue, the cruise industry has actively embraced the initiative of "green shipping". This study investigates the relationship between ESG and corporate financial performance (CFP) in the global cruise sector. This paper utilizes the sample data from the world's largest cruise companies over 2012-2023, to examine the ESG-CFP relationship by a regression model. The results indicate that ESG practices in cruise companies negatively influence CFP, which is further impacted by financial constraints. Furthermore, the heterogeneity analysis suggests that the high time interest earned (TIE) ratios and low total annual greenhouse gas (GHG) emissions worsen the adverse impacts of ESG on CFP. These findings contribute to the theoretical research on ESG and provide practical guidance for cruise industry operators and investors in their decision-making.

## 1. Introduction

As a component of the shipping and tourism sector, the cruise industry is characterized by its high level of globalization and capital intensity. Over the recent decades, it has been developing investment plans that have yielded substantial gains. Notably, in 2018, the industry contributed $134 billion to the global GDP [1], which highlights its positive influence on the economic advancement and its active participation in the global economy [2, 3].

Nonetheless, in recent years, the cruise industry has suffered financial distress as a consequence of global uncertainties, such as COVID-19 pandemic and geopolitical conflicts like the Palestinian-Israeli conflict, which are referred to as "black swan" events [4]. The economic contribution of global cruise industry decreased by 59% between 2019 and 2020, from $154.5 billion to $63.4 billion [5]. Although a few cruise corporations have been gradually re-starting business operations, the entire sector as a whole remains unrecovered. Consequently, cruise corporations are prioritizing immediate profitability and enhancements in operational efficiency. Meanwhile, given the trend of sustainable development, they are compelled to concentrate on ESG and conform with international regulations. Stakeholders increasingly prioritize ESG factors when making investment and consumption decisions [6]. Therefore, it's critical to balance the corporate financial success with ESG considerations. However, relevant studies in this area remain limited. To improve competitiveness and sustainable expansion of cruise firms, it is vital to explore the relationship between ESG and CFP.

ESG was initially proposed in the 2006 United Nations Principles for Responsible Investment (UNPRI) report [7]. This paradigm highlights the importance of corporate commitment to social and environmental responsibility and improvements of relationships with stakeholders such as suppliers, consumers, and shareholders [8]. ESG urges firms to shift from a profit-driven strategy to one that promotes a variety of goals, including environmental preservation and social responsibility [9]. As a result, cruise companies are emphasizing on ESG measures. For example, investors in the cruise and transportation industries place greater emphasis on ESG evaluations when making decisions. According to the "Rate the Raters 2023" poll, 65% of investors include ESG ratings as a critical factor in evaluating a company's performance.

However, it is crucial to assess whether a corporate's improvement in ESG has a positive effect on CFP. This evaluation is essential in determining whether company's social value and economic value are mutually reinforcing, thereby realizing the sustainable development [10]. In recent years, despite a growing number of academic study on the influence of ESG performance on financial returns [11,12,13], disagreement does exist in the views concerning the relationship between ESG and CFP [14]. Multiple studies have shown a beneficial relationship between ESG and CFP, as the ESG improves CFP by lowering financing costs and nurturing stakeholder confidence in the organization [15, 16]. However, it is also argued that engaging in ESG activities might potentially harm a company's cash flow, leading to a decrease in CFP [17]. Furthermore, some researches indicate that the significant relationship between ESG performance and CFP does not necessarily exist. In a word, arguments are different on the relationship between ESG and CFP.

In the cruise industry, financial constraints significantly influence CFP because of the enormous financial transactions. Financial constraints refer to the limitations on a company's ability to obtain cash flows, often cited as a primary impediment to the long-run growth of cruise corporations. Given the financial distress, cruise firms tend to depend more on internal financing, which increases the expense associated with obtaining external funding. Furthermore, financial constraints may restrict a company's capacity to allocate funds for investments and development, negatively impacting its overall financial performance in the long-run [16]. These constraints also affect ESG. The cruise companies with better financial flexibility may better allocate funds to ESG initiatives, which can improve ESG performance. In contrast, the greater financial restrictions on companies, the more likely they are to prioritize preserving current revenues, which results in fewer involvements in ESG projects. Moreover, previous studies imply that improving ESG would not only increase stakeholder trust in the company's prospects, but will also reduce financing costs, thereby improving CFP [18].

Despite the studies on the impact of ESG on CFP, the academic gaps still remain in the shipping and cruise industry. Moreover, previous researches have explored the association between financial constraints and financial performance in the cruise industry, without considering ESG. In addition, given the unique characteristics of the cruise industry, factors such as TIE ratios and total GHG emissions are critical for evaluating ESG and CFP. TIE ratios provide useful information about enterprises' financial health and capacity to repay loans [19]. While total GHG emissions demonstrate their commitment to global decarbonization [20]. Nevertheless, the study in this field is currently limited.

Given the current limitations, this paper contributes as follows:

First, this study focuses on the relationship between ESG and CFP in the cruise sector, using the data from aforementioned companies (i.e., Carnival Corporation (CCL), Royal Caribbean Cruises (RCL), and Norwegian Cruise Line Holdings (NCLH)[1]) over 2012-2023, as case studies. This study is ground-breaking in terms of sustainable development and financial strategy within the cruise industry. Moreover, it also delves into novel perspectives on the development of "green shipping".

In addition, this paper employs heterogeneity analysis to evaluate the impacts of industry-specific factors (i.e., the TIE ratio and total GHG emissions) on ESG and CFP. The TIE ratio is an important indicator of cruise companies' financial strategy and stability, with a substantial influence on its CFP. Moreover, in response to growing concerns about GHG emissions from the shipping industry, cruise companies are introducing carbon-cutting strategies, which significantly impacts the assessments of ESG performance. In other word, this paper presents a novel methodology.

Furthermore, the cruise corporations pay more attention to sustainability in response to the financial and operational issues result from increasing global uncertainties. The research focuses on how financial constraints may impact the correlation between ESG and CFP in the cruise industry. Particularly, it evaluates the current restraints on corporate finance and advises thorough analysis on the implications of these constraints.

The subsequent portions of the research are structured as follows: Section 2 examines the pretinent literature concerning the correlation between ESG performance, CFP, and the mitigating influence of financial constraints, which is followed by section 3 that provides a thorough analysis of the mechanisms and develops hypotheses. While section 4 provides an overview of the data sources, terminology, and methodology employed in constructing the model. Section 5 offers a comprehensive evaluation of the collected data, followed by a discourse on the empirical findings. Finally, Section 6 encompasses the study's conclusions, recommendations, potential avenues for future research, and a discussion on limitations inherent in the study.

---

[1] The three cruise companies were selected based on their combined supremacy in the global cruise industry, holding approximately 80% of the market share. The sample of large cruise operators represents a significant portion of the business,since they carry 78% of passengers and create 75.3% of revenues.

## 2. Literature Review

### 2.1. ESG Performance and CFP

Prior research has delved into the connection between ESG performance and CFP, suggesting that the financial success is consistently correlated with ESG variables [21]. For instance, Rodríguez-Fernández et al. (2019) [22] identified a positive correlation between ESG and CFP, as measured by Return on Assets (ROA) and Return on Equity (ROE). Moreover, Zhao et al. (2018) [23] demonstrated that ESG performance in the energy and power sector may increase company profits. However, there is not a unanimous global perspective on the link between ESG and CFP. For example, Auer and Schuhmacher (2016) [24] examined this relationship in different regions and discovered that investment-focused strategies enhance ESG performance for investors in the Asia-Pacific and the US, while the case in Europe is on the opposite side.

Nevertheless, current research on ESG and CFP remains conflicting and inconclusive. The Optimal Social Responsibility Hypothesis proposes that the relationship between profitable companies and Corporate Social Responsibility (CSR) may follow an inverted U-shaped pattern [14], which indicates that the CFP is negatively affected at the early stage of introduction of ESG practice, since the initial costs is larger than the profits. While the CFP starts to grow when a corporate's ESG standards achieve a certain level.

Given the previous findings that ESG is either negatively or neutrally correlated with CFP [25]. Neoclassical researchers believe that enhancing ESG performance may increase corporate expenditures and reduce cash flow [26, 27]. The CFP is likely to be worse off if excessive or inappropriate investments in ESG initiatives [17]. For example, excessive investments in emission reduction or natural resource enhancement can lead to excessive costs that negatively affect the financial performance [28, 29]. Additionally, the growth of cash flow, staff, and infrastructure may affect a company's finances, particularly if it accepts cost leadership strategy [30]. Research has also indicated that a negative correlation between ESG and CFP tends to be in countries with greater self-confidence and gender equality [31]. Furthermore, Friede et al. (2015) [15] found that the negative correlation between ESG and CFP is stronger in European and Asian/Australian developed markets compared to North American companies.

### 2.2. Financial Constraints, ESG Performance and CFP

Cruise ship operators may face financing difficulties due to financial constraints, since the financial limitations make it more difficult for a corporation to obtain external funding, leading to higher financing costs and reduced internal cash flow [32]. As financial restrictions grow, firms might resort to aggressive financing tactics, which can increase uncertainties and share price volatility that raise the overall risk of company [33, 34]. This increased risk exposure lowers corporate value and negatively affects the CFP.

Studies have also examined the relationship between ESG and financial constraints [35]. Luo et al. (2023) [36] and Feng et al. (2024) [37] identified a positive association between robust ESG practices and better access to credit financing for businesses. Companies with poor ESG may face higher financing costs, budgetary constraints, and information misalignment issues [38]. In contrast, firms with strong ESG performance typically present improved transparency for investors, which is helpful to mitigate risks associated with government policies [39]. Strong ESG performance can not only bolster a company's standing, enhance public perception, and build trust with stakeholders, but also reduces decision-making risks for creditors and investors by addressing information imbalance. Therefore, companies are more inclined to introduce additional investments from capital markets and financial institutions, thereby reducing their financial constraints [21, 40].

The relationship between ESG performance and CFP have been better understood after literature review. Nevertheless, spaces remain for further explorations on details.

Existing studies examing the correlation between ESG and CFP is inconclusive and ambiguous for following reasons. (1) Capital investment strategies vary across companies and markets [41]. (2) ESG disclosure and scoring criteria vary by country and region, making it difficult to compare study results [42]. (3) Different levels of development in emerging and established markets may impact ESG performance, as stakeholder support for ESG initiatives varies [43].

Furthermore, limitations exist in the research samples and views regarding the examination of ESG performance, CFP, and budgetary constraints. Domestic and international studies show that the ESG data is hard to access as various organizations evaluate ESG scores differently. Moreover, the research focusing on the shipping and cruise sectors is limited. Despite studies investigating the three sub-components limited study examines the environmental implications of ESG, particularly the impact of carbon dioxide emissions. Considering the concerns raised by cruise ship carbon emissions, the limited number of studies in this area is unexpected.

This study aims to fill a scholarly hole in the existing literature. By using regression models to analyze cruise companies' ESG and CFP performance, this study contributes to the current understanding of their relationship. Additionally, a moderating impact model is employed to evaluate "ESG performance-financing constraints-CFP", further exploring the determinants of the link between ESG and CFP. The research also includes heterogeneity analysis on TIE ratios and total GHG emissions, with considering cruise enterprises' unique characteristics. This analysis aims to better understand how these two indicators affect ESG performance, CFP, and financial constraints. These discoveries contribute to the literature on the current body knowledge on the factors that influence the financial success of cruise companies.

## 3. Hypothesis

### 3.1. The Relationship between Cruise Companies' ESG Performance and CFP

The sector-specific factors should be considered when examining ESG performance and financial outcomes in cruise enterprises. Cruise companies involve significant financial investments and environmental concerns. Therefore, exogenous uncertainties such as the COVID-19 pandemic and carbon emission reduction affect their financial performance and investment strategies.

Cruise corporates have substantial environmental challenges since they represent one of the most energy-intensive sectors in the tourist industry [44]. To address the issue, cruise companies are striving to balance between economic development and energy conservation. In recent years, cruise companies have significantly invested in reducing GHG emission. Nevertheless, the financial performance impacts of these environmental initiatives and resource reallocations are not immediate and may initially cause a drop in financial indicators [45].

The COVID-19 pandemic introduced unprecedented challenges to the cruise industry, including a considerable fall in passenger demand, and increased costs associated with outbreak prevention, which led to a significant drop in profitability. Meanwhile, because of social demand of ESG, cruise firms have been obliged to participate in more ESG initiatives. These investments may negatively affect the financial condition of cruise companies in the short-run.

Based on the above explanation, the following hypothesis is suggested:
H1: ESG performance of cruise companies has a detrimental effect on CFP.

### 3.2. The Impact of Financial Constraints on Cruise's ESG performance and CFP

Financial constraints raise a significant challenge for cruise companies, especially in light of current global economic uncertainty and volatility in financial markets. Cruise companies commonly accept debt-financing instruments including bank loans and bond issuances for capital-intensive initiatives [46].

These financial constraints influence cruise firms' decisions. The stricter restrictions on firms, the more difficult for firms to get external funding. Consequently, they might resort to more costly financing options such as high-interest loans, which potentially restricts investment capacity and erodes CFP. Given the financial crisis, insufficient expenditures on marketing and brand development impede cruise companies' ability to maintain high service standards, which makes the company's market competitiveness worse off. Enterprises may struggle to get adequate funding for ESG, which may negatively influence ESG performance.

Furthermore, funding restrictions have limited strategic decision-making in cruise companies, which leads them to take more conservative strategic approaches. The conservatism may cause the preference for established industries and items over riskier projects like ESG investments. Consequently, lower expenditures on ESG may affect ESG performance of cruise companies.

Thus, the following hypothesis is raised:
H2: Financial constraints of cruise companies have a detrimental moderating effect on ESG and CFP.

## 4. Data Description and Methodology

### 4.1. Sample Selection and Data Sources

As cruise firms have been scrutinized by financial market, this study examines a sample of three cruise enterprises specifically listed on the New York Stock Exchange. Compared to other segments in the travel industry, the cruise market is less fragmented, since three particular cruise firms dominate the cruise industry, by serving 78% of passengers and generating 75.3% of revenues. To be specific, as illustrated in Figures 1 and 2, CCL is the market leader with the biggest proportion of passengers (42.9%) and highest revenue share (37.3%). It operates eight cruise brands, including P&O Cruises, Princess, and AIDA. RCL accounts for 25.7% in passenger services and earns 23.9% of total revenue. It covers three cruise brands: Royal Caribbean, Celebrity, and Silversea. While NCLH accounts for 9.4% of passenger services and generates 14.1% of total revenues, with three cruise brands: Norwegian, Regent Seven Seas, and Oceania Cruises. Moreover, these firms are listed and their shares are traded on New York and London stock markets, providing empirical insights on their stock market behaviour, performance, and capitalization [19]. Therefore, these three dominant cruise lines are considered to be representative and instructive for assessment based on uniform criteria.

Figure 1. Revenue Distribution of Global Cruise Companies in 2023.

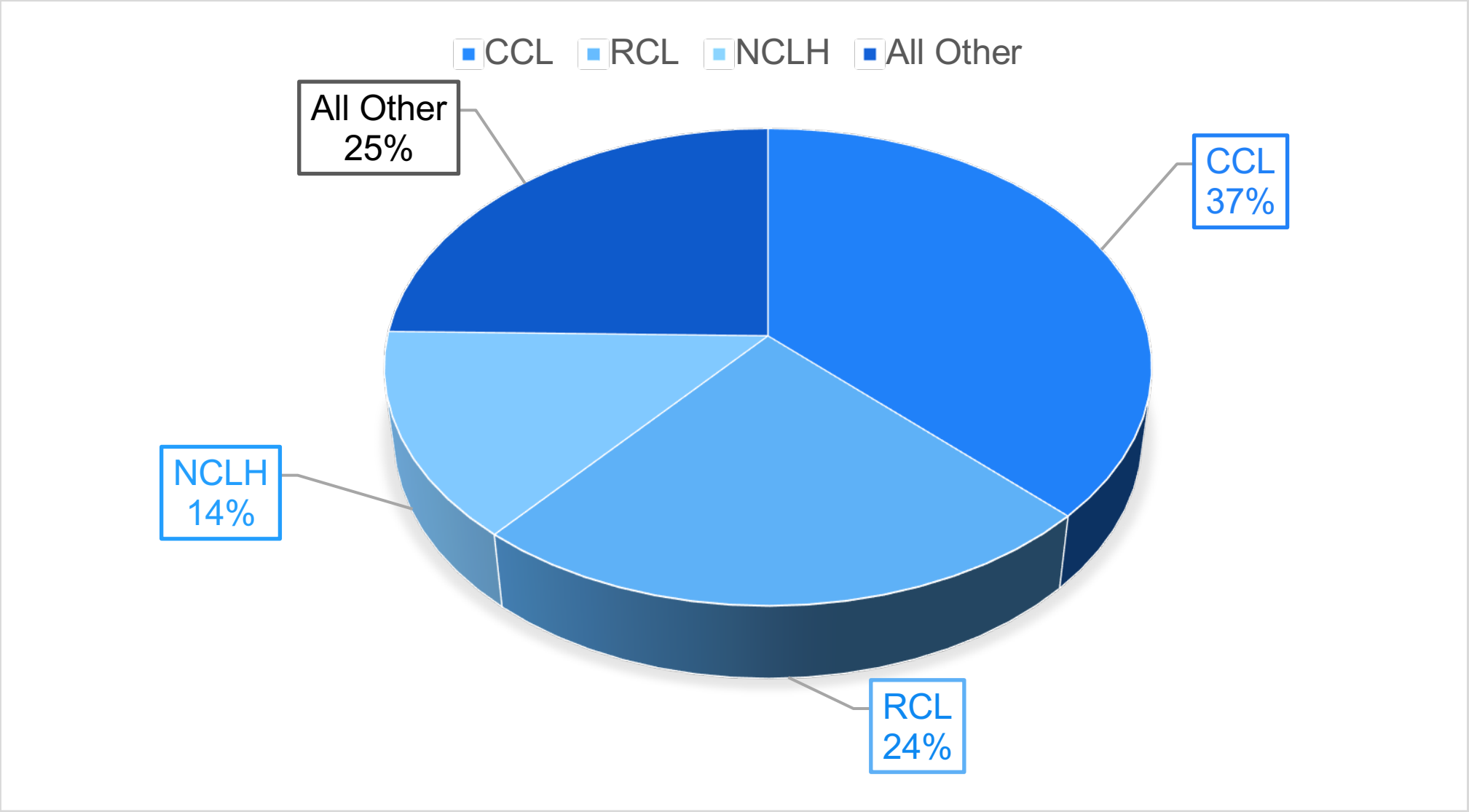

Source: https://cruisemarketwatch.com/market-share/#/

Figure 2. Distribution of passengers by global cruise companies in 2023.

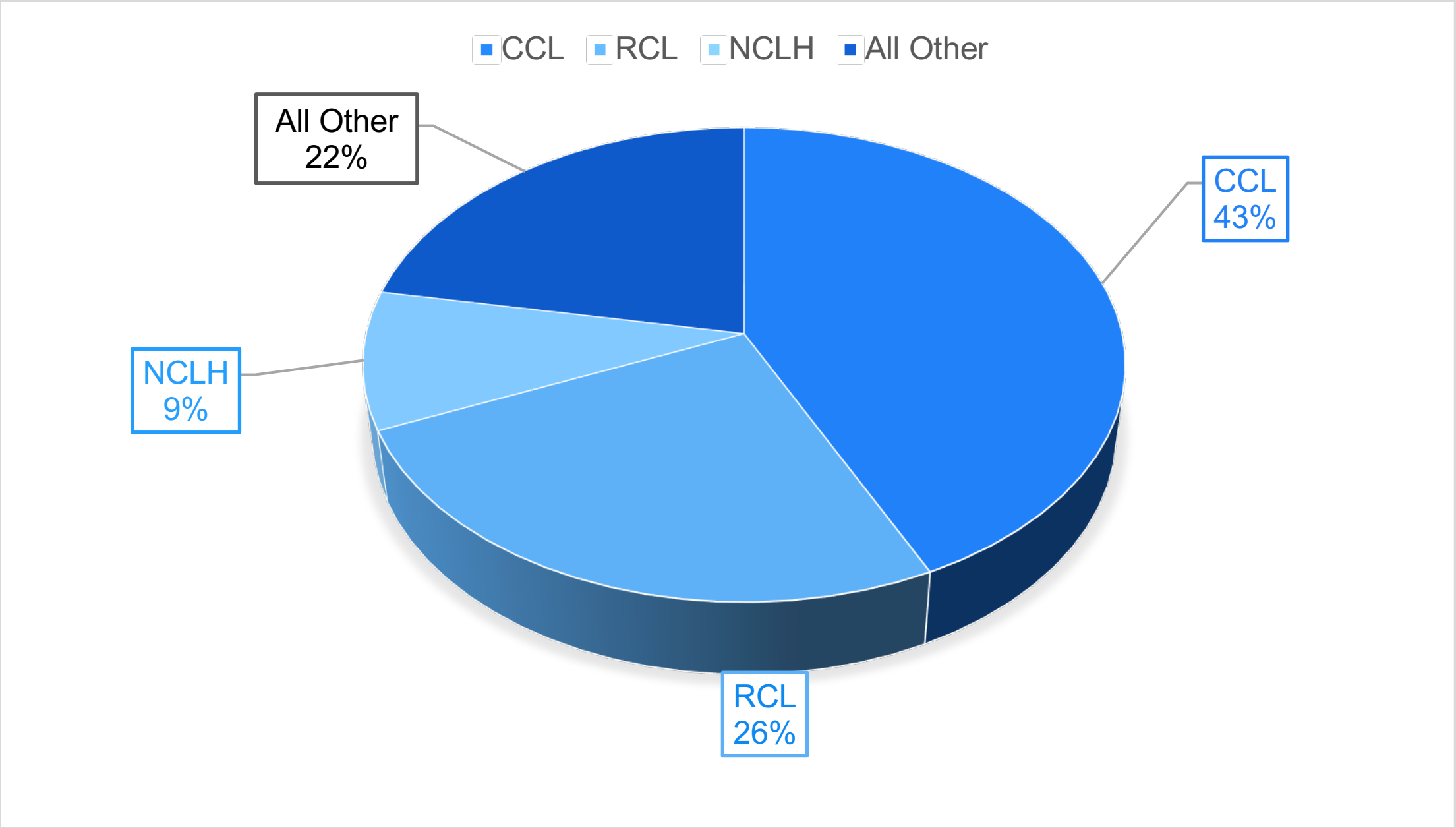

Source: https://cruisemarketwatch.com/market-share/#/

The sample period for this research is 2012-2023, it was chosen based on the availability of data and the pandemic. ESG data was obtained from Bloomberg, while the CFP data was from Bloomberg, Wind and cruise business annual reports. Then, the CFP and ESG statistics were matched for analysis. SPSS and STATA 16.0 were used to conduct empirical analyses of the final dataset after eliminating enterprises with special treatment (ST) status and those with missing data throughout the sample period.

## 4.2. Variable Definitions

4.2.1. Dependent Variable

Based on previous research, ROA has been recognized as the standard measure for assessing CFP [9, 47]. ROA measures a company's ability to create net profit by effectively using its assets. It reflects the company's skill in managing both fixed and current assets [19].

Figure 3. ROA dynamics of three leading cruise companies over 2012- 2023.

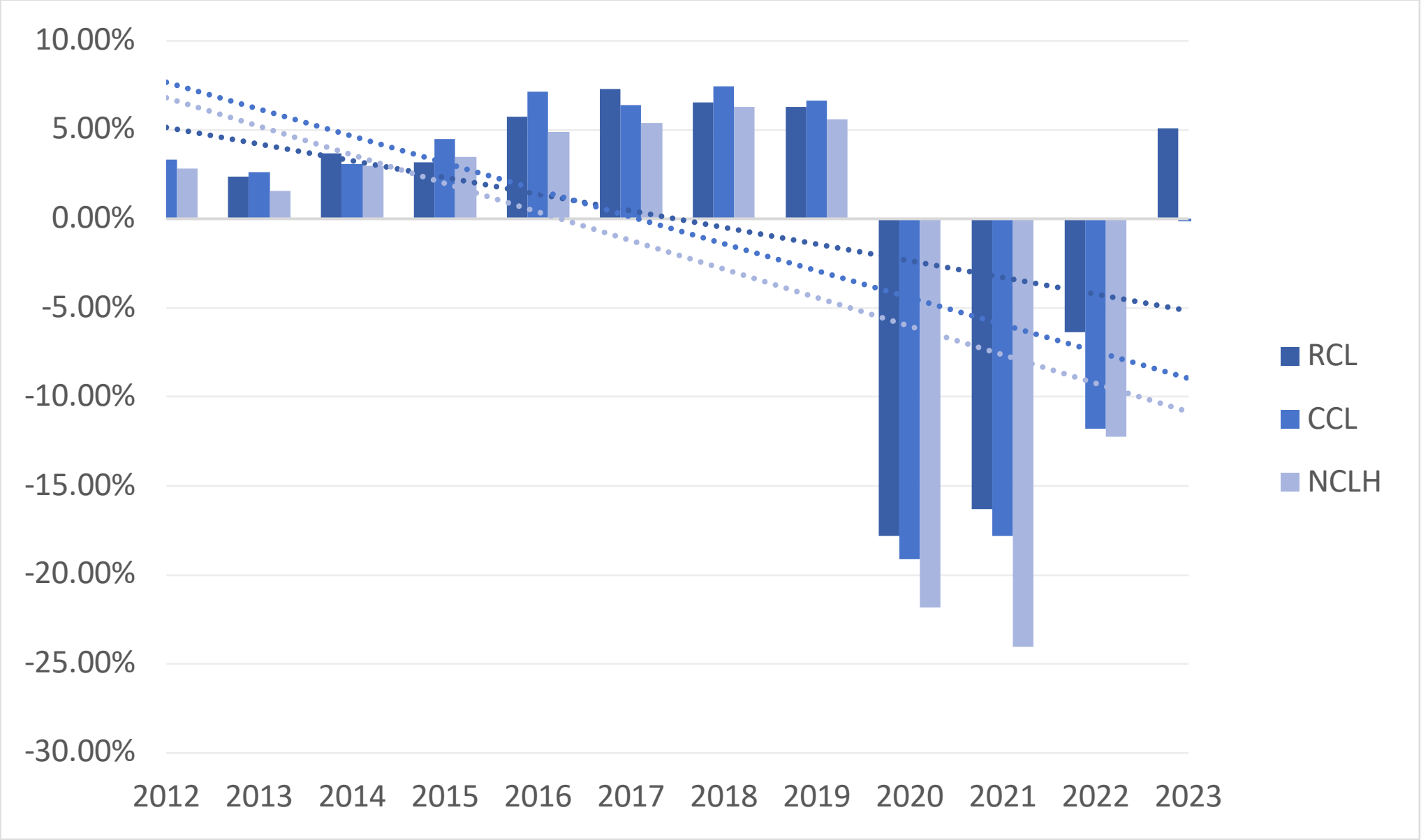

4.2.2. Independent Variable

ESG performance. This research use the ESG Financial Materiality Scores sourced from the Bloomberg database to evaluate ESG. A higher score signifies a superior ESG performance.

4.2.3. Moderating Variable

Financing constraints (KZ) are moderators in the relationship between independent and dependent variables. Prior research has explored the moderating influence of financial limitations when analyzing the relationship between ESG and CFP [38]. This research adopts the KZ index proposed by Bai et al. (2022) [32], to assess a firm-level financing constraints. The KZ index is calculated through an ordered logistic regression model, utilizing the firm's operating cash flow, gearing ratio, cash holdings and Tobin's Q as independent variables. A higher KZ index indicates greater financial distress and more pronounced financing constraints for the firm.

4.2.4. Control Variables

Drawing insights from previous studies [40, 48], this study incorporates financial leverage (LEV), firm size (SIZE), asset structure (AS), cash ratio (CASH), and earnings per share (EPS) as control variables. The precise definitions of them are provided in Table 1.

**Table 1.** The definition of variables.

| Variable | Variable name | Variable | Variable measures |
|---|---|---|---|
| Dependent variable | Corporate Financial Performance | CFP | Net Profit/Total assets |
| Independent variable | ESG performance | ESG | ESG Financial Materiality Scores |
| Mediators | Financial constraints | KZ | The KZ index is calculated through an ordered logistic regression model utilizing the firm's operating cash flow, gearing ratio, cash holdings and Tobin's Q. |
| Control variables | Financial leverage | LEV | Total liabilities/Total assets |
| | Enterprise scale | SIZE | Natural logarithm of total assets for the year |
| | Asset structure | AS | Fixed assets/Total assets |
| | Cash ratio | CASH | Net cash flow from operating activities/Total assets |
| | Earnings per share | EPS | Net profit after tax/Total equity |

4.3. Model Construction

This research utilizes regression Model (1) to test Hypothesis 1 (H1): ESG performance has a detrimental impact on the CFP. The specific model is set as follows:

$$CFP_{i,t} = \alpha_0 + \alpha_1 ESG_{i.t} + \alpha_2 KZ_{i,t} + \alpha_3 LEV_{i,t} + \alpha_4 SIZE_{i,t} + \alpha_5 AS_{i,t} + \alpha_6 CASH_{i,t} + \alpha_7 EPS_{i,t} + \varepsilon_{i,t} \quad (1)$$

where, α denotes the constant value, $\alpha_1 \sim \alpha_n$ represents the factor associated with the specific control variable, ε represents the random error term, i is for the business in the equation above, and t indicates the year.

This research establishes regression Model (2) to test Hypothesis 2 (H2): financial constraints demonstrate a negative moderating effect between ESG and CFP. The specific model is as follows:

$$CFP_{i,t} = \beta_0 + \beta_1 ESG_{i.t} + \beta_2 KZ_{i,t} + \beta_3 ESG_{i.t} * KZ_{i,t} + \beta_4 LEV_{i,t} + \beta_5 SIZE_{i,t} + \beta_6 AS_{i,t} + \beta_7 CASH_{i,t} + \beta_8 EPS_{i,t} + \delta_{i,t} \quad (2)$$

where, β denotes the constant value, $\beta_1 \sim \beta_n$ represents the factor associated with the specific control variable, δ represents the random error term, i is for the business in the equation above, and t indicates the year.

## 5. Results and Discussion

### 5.1. Descriptive Statistics

Table 2 displays the descriptive statistics for variables. The mean value of CFP is greater than the median value, indicating a right-skewed distribution in financial performance. Furthermore, the conspicuous disparity between the highest and lowest figures implies significant fluctuations in CFP. The ESG scores ranging from 1 to 9 with an average of 4.113 indicates a notably low level of ESG performance, highlighting the spaces for improvement in this area. The mean and median values of KZ are at 2.513 and 2.535, respectively, indicating that the leading cruise companies face similar constraints in corporate financing. The consistency may be ascribed to the influence of the pandemic and the persistent constraints on fundraising methods. Analysis of the control variables shows that the variables are generally concentrated and stable, with less vulnerability to sudden changes.

**Table 2.** Descriptive statistics for variables.

| Variable | Maximum | Minimum | Median | Mean |
|---|---|---|---|---|
| CFP | 0.074 | -0.241 | 0.031 | -0.009 |
| ESG | 5.77 | 1.31 | 4.52 | 4.113 |
| KZ | 2.705 | 2.213 | 2.535 | 2.513 |
| LEV | 0.996 | 0.386 | 0.609 | 0.641 |
| SIZE | 10.889 | 8.689 | 10.122 | 10.118 |
| AS | 0.904 | 0.71 | 0.821 | 0.815 |
| CASH | 1.72 | 0.02 | 0.06 | 0.219 |
| EPS | 8.95 | -27.05 | 1.765 | -0.776 |

### 5.2. Benchmark Effects Regression Analysis

Model (1) in Table 3 shows the findings of the benchmark regression analysis assessing the association between ESG and CFP. All regressions include industry and year-fixed effects. The regression coefficients exhibit statistical significance at the 1% level, indicating a negative relationship between ESG and CFP. Specifically, a single unit rise in ESG is linked to a 1% reduction in CFP.

This negative association between ESG and CFP is likely influenced by the unique characteristics of the cruise industry, which is vulnerable to exogenous economic factors and seasonal fluctuations, leading to unstable cash flows. Factors such as the global epidemic and decarbonization efforts have further impacted cruise companies' earnings and cash flow.

Investments in ESG initiatives, such as emission reduction strategies, can be costly and may not yield immediate returns, negatively affecting the cash flow position of cruise firms.

Therefore, the conclusion supports H1 and aligns with previous research [49].

5.3. Endogeneity Tests

To slove potential endogeneity from reverse causality, the correlation between ESG and CFP is investigated by using one-period and two-period lag length for the explanatory variable CFP. Table 3 presents the regression outcomes for Model (2) and Model (3). In Model (2), the regression result for one-period lagged CFP is -0.009 and for two-period lagged CFP is -0.007, with evident negative coefficients at the 1% level. The results further support H1.

The comparisons of the current effect observed in Model (1) reveal that ESG has a more noticeable negative influence on CFP in the short-term. However, the positive impacts of ESG investments may gradually manifest over time, resulting in a reduction in CFP, which implies that the influence of ESG factors on CFP is not immediate and requires the accumulation of technological advancements and time before reflecting in the financial performance of enterprises.

5.4. Robustness Tests

In all regression analyses, the model is reorganized by substituting ROA with ROI, which is a metric for CFP, while considering firm-specific characteristics, industry factors, and temporal effects. ROI is an important indicator of operational profitability, as it measures the total capital investment and emphasizes core operational efficiency.

The ESG coefficient in Model (4) shown in Table 3 is -0.01, which is statistically significant at the 1% level. This implies that an enhancement in ESG leads to a decline of nearly 1% on average in CFP. This supports H1 and remain robust under statistical scrutiny at the 5% significance level, therefore confirming the validity of the findings.

**Table 3.** Regression findings of the relationship between ESG performance and CFP.

| Variable | Model（1） | Model（2） | Model（3） | Model（4） |
|---|---|---|---|---|
| | | 1st stage | 2nd stage | Replaced dependent variable |
| | | $ROA_{i,t}$ | | $ROI_{i,t}$ |
| ESG | -0.01**<br>(-2.66) | -0.009***<br>(-2.757) | -0.007**<br>(-2.161) | -0.01**<br>(-2.66) |
| LEV | -0.115***<br>(-4.282) | -0.114***<br>(-4.576) | -0.11***<br>(-4.542) | -0.115***<br>( -4.282) |
| SIZE | 0.008<br>(0.886) | 0.007<br>(0.385) | 0.002<br>(0.328) | 0.008<br>(0.886) |
| AS | 0.36***<br>(3.662) | 0.352***<br>(3.816) | 0.242***<br>(3.13) | 0.36***<br>(3.662) |
| CASH | -0.071***<br>(-4.799) | -0.071***<br>(-5.088) | -0.076***<br>(-5.651) | -0.071***<br>(-4.799) |
| EPS | 0.005***<br>(7.776) | 0.005***<br>(8.212) | 0.005***<br>(8.678) | 0.005***<br>(7.776) |
| Industry | Yes | Yes | Yes | Yes |
| Year | Yes | Yes | Yes | Yes |
| Constant | -0.245**<br>(-2.062) | -0.231**<br>(-2.164) | -0.109<br>(-1.169) | -0.245**<br>(-2.062) |
| R-squared | 0.92 | 0.92 | 0.914 | 0.92 |

Note: Values included in parenthesis represent the robust standard errors; ***, **, and * reflect 1%, 5%, and 10% significance levels, respectively.

(1) Comparative regression findings regarding the effects of ESG on CFP.

(2) Regression results illustrating the effects of ESG on CFP following a lagged treatment over one period.

(3) Regression findings on the correlation between ESG and CFP impact subsequent to a lag in the second treatment period.

(4) Regression results obtained by substituting ROA with ROI as an indicator of CFP.

## 5.5. Analysis of Moderating Variables

Model (1) in Table 4 investigates the correlation between CFP and KZ. The regression analysis shows a statistically evident coefficient of -0.17 for financial constraints at 1% level. It demonstrates that financial constraints negatively impact CFP. The regression model in Model (2) include the interaction term between ESG and KZ. The result indicates that the ESG performance coefficient is -0.005, a statistically finding at the 1% significance level. While the interaction term, ESG×KZ, has a coefficient of -0.003, which is statistically significant at the 1% level. These findings indicate that financial limitations have a detrimental moderating impact. As a result, the analytical results confirm the validity of H2.

Since the COVID-19 has raised issues for cruise operators, the requirement for external financing to enhance cash reserves through the issuance of long-term debt has considerably risen [50, 51]. However, the growing reliance on external financing has led to financing constraints that hinder the survival and growth of companies. Given the funding constraints, companies face financial distress, reduced financial performance, and the possibility of reducing proactive ESG initiatives to maintain current output. These findings align with the previous studies [51, 52].

**Table 4**. Regression findings of the relationship between ESG performance, CFP and financing constraints.

| | Model（1） | Model（2） |
|---|---|---|
| Variable | $ROA_{i,t}$ | |
| $ESG_{i,t}$ *$KZ_{i,t}$ | | -0.003***<br>(-3.315) |
| $KZ_{i,t}$ | -0.17***<br>(-5.276) | -0.16***<br>(-5.031) |
| ESG | -0.009***<br>(-2.927) | -0.005**<br>(-2.592) |
| LEV | -0.204***<br>(-9.38) | -0.205***<br>(-9.355) |
| SIZE | 0.000<br>(-0.049) | 0.003<br>(0.454) |
| AS | 0.367***<br>(4.577) | 0.384***<br>(4.718) |
| CASH | -0.073***<br>(-6.117) | -0.073***<br>(-6.068) |
| EPS | 0.004***<br>(7.794) | 0.004***<br>(7.708) |
| Industry | Yes | Yes |
| Year | Yes | Yes |
| Constant | 0.311**<br>(2.546) | 0.244*<br>(1.943) |
| R-squared | 0.95 | 0.951 |

Note: Values included in parenthesis represent the robust standard errors; ***, **, and * represent 1%, 5%, and 10% significance levels, respectively.

(1) Regression results for the relationship between CFP and KZ.

(2) Regression results with the inclusion of the interaction term between ESG performance and KZ.

## 5.6. Analysis of Heterogeneity

### 5.6.1. Heterogeneity Analysis under Different TIE Ratios

The times interest earned (TIE) ratio measures the correlation between a company's net operating income and its annual interest expense obligations [19]. It is used to assess a company's ability to fulfil its long-term financial obligations. A higher TIE suggests that the company has sufficient funds to continue its operations after servicing its debt.

This study categorizes the sample into groups with low and high TIE ratios for separate analyses, based on the mean value, to evaluate the influence of financial limitations on ESG and CFP. Table 5 displays the precise regression findings for the given grouping.

1. Model (1) and Model (3) show that in high TIE ratios group, the coefficient of ESG is -0.012 and it is statistically significant at 1% level, whereas the low TIE ratio group shows no significant results. High TIEs indicate that cruise firms can repay more loans and are more likely to maintain continuity of operations during the economic fluctuations. Therefore, they have more resources and time to invest in ESG projects. However, due to cruise industry characteristics, ESG still negatively affects CFP, as ESG investments do not yield immediate returns. In contrast, companies with low TIE ratio prioritize investments in fleet and market growth over ESG. ESG and CFP are less associated in this group.

2. Model (2) and Model (4) illustrate that financial constraints hinder the cash flow of both high and low TIE ratio companies. In high TIE ratios group', this adverse influence is more apparent and statistically significant at 1%. These companies are more financially stable, making them more attractive to banks and investors. Their financial liquidity is well-prepared to pay higher financing costs. Additionally, they are able to raise more capital and finance for more projects. Nevertheless, the negative impact on ESG performance becomes more notable once their financing is constrained. In contrast, companies with low TIE ratios may be unable to borrow additional funds, leading to more cautious operations. Since they need less funding, the financing constraints have less impact on their CFP and ESG.

**Table 5.** Regression results under heterogeneity of TIE.

| | Low TIE ratios | | High TIE ratios | |
|---|---|---|---|---|
| | Model（1） | Model（2） | Model（3） | Model（4） |
| Variable | | $ROA_{i,t}$ | | |
| $ESG_{i,t}$ *$KZ_{i,t}$ | | -0.003 | | -0.002*** |
| | | (-1.496) | | (-3.406) |
| $KZ_{i,t}$ | | -0.115* | | -0.184*** |
| | | (-2.42) | | (-4.395) |
| ESG | -0.006 | -0.002 | -0.012*** | -0.006** |
| | (-1.055) | (-0.506) | (-3.287) | (-3.002) |
| LEV | -0.028 | -0.113* | -0.242*** | -0.286*** |
| | (-0.595) | (-2.45) | (-6.947) | (-9.711) |
| SIZE | 0.003 | 0.002 | -0.019* | -0.008 |
| | (0.255) | (0.194) | (-1.973) | (-0.904) |
| AS | 0.616*** | 0.684*** | 0.197** | 0.225** |
| | (4.355) | (4.05) | (2.225) | (2.927) |
| CASH | -0.052** | -0.05* | -0.032** | -0.048*** |
| | (-2.862) | (-2.215) | (-2.499) | (-4.92) |
| EPS | 0.004*** | 0.003** | 0.008*** | 0.006*** |
| | (4.712) | (3.918) | (9.96) | (10.971) |
| Industry | Yes | Yes | Yes | Yes |
| Year | Yes | Yes | Yes | Yes |
| Constant | -0.501** | -0.202 | 0.227* | 0.595*** |
| | (-2.831) | (-1.003) | (1.936) | (4.1) |
| R-squared | 0.94 | 0.959 | 0.963 | 0.977 |

Note: Values included in parenthesis represent the robust standard errors; ***, **, and * represent 1%, 5%, and 10% significance levels, respectively.
(1) Comparative regression findings regarding the influence of low TIE ratio groups' ESG performance on CFP.
(2) Regression results after introducing the interaction term between low TIE ratio groups' ESG performance and KZ.
(3) Comparative regression findings regarding the influence of high TIE ratio groups' ESG performance on CFP.
(4) Regression results after introducing the interaction term between high TIE ratio groups' ESG performance and KZ.

5.6.2. Heterogeneity Analysis under Different Greenhouse Gas (GHG) Emission Totals

Carbon emission reduction has been a major issue in the shipping industry. The International Maritime Organization (IMO)'s "Initial Strategy on Reducing GHG Emissions from Shipping" aims to cut GHG emissions by 50% by 2050, compared to 2008 levels. Total GHG emissions are also crucial to measure the environmental element (E) of ESG.

To investigate carbon emissions from different fuel sources, emissions must be converted into CO2 equivalents. This study examines GHG Scope 1 and GHG Scope 2 Location-Based emissions in Bloomberg's Environmental section of these cruise companies. These two categories are combined to compute total GHG emissions, which are then divided into low and high emission groups by average values for further analysis.

1. Model (1) and Model (3) in Table 6 present a larger adverse correlation between ESG and CFP in companies with low total GHG emissions. This could be because cruise firms with lower overall GHG emissions invest more in emission reduction. Thus, they have a more substantial influence on ESG. However, cost-intensive initiatives like investing in LNG-powered ships and On-shore Power Supply require significant initial expenditure. These investments do not yield immediate profits but are designed to reduce emissions. Consequently, the zero-profit affects CFP with a coefficient of -0.045. This outcome further supports the theoretical explanations of the negative relationship between cruise companies' ESG and CFP in the benchmark regression results.

2. Model (2) and Model (4) in Table 6 demonstrate that financial constraints of cruise companies with lower overall GHG emissions have a stronger inverse moderating effect. When these corporations invest more in emission reduction and raising environmental awareness, their market status, influence, and brand value improvements. Consequently, their performance in ESG assessment has become more significant.

**Table 6.** Regression results under heterogeneity of GHG emission totals.

| | Low total GHG emissions | | High total GHG emissions | |
|---|---|---|---|---|
| | Model（1） | Model（2） | Model（3） | Model（4） |
| Variable | | $ROA_{i,t}$ | | |
| $ESG_{i,t}$ $*KZ_{i,t}$ | | -0.004** | | -0.002 |
| | | (-2.897) | | (0.288) |
| $KZ_{i,t}$ | | -0.28*** | | -0.088** |
| | | (-4.506) | | (-2.864) |
| ESG | -0.045* | -0.008* | 0.001 | 0.006 |
| | (-2.224) | (-2.23) | (0.072) | (0.93) |
| LEV | -0.157* | -0.308*** | -0.079*** | -0.136*** |
| | (-1.912) | (-6.299) | (-3.775) | (-6.152) |
| SIZE | 0.111* | 0.012 | 0.031** | 0.027* |
| | (2.101) | (0.749) | (2.461) | (2.167) |
| AS | 0.749* | 0.402** | 0.405*** | 0.434*** |
| | (2.298) | (2.459) | (4.936) | (4.654) |
| CASH | -0.033 | -0.059** | -0.09*** | -0.095*** |
| | (-0.94) | (-3.541) | (-5.967) | (-5.8) |
| EPS | 0.007*** | 0.003*** | 0.005*** | 0.005*** |
| | (4.393) | (4.209) | (6.969) | (6.138) |
| Industry | Yes | Yes | Yes | Yes |
| Year | Yes | Yes | Yes | Yes |
| Constant | -1.414* | 0.539* | -0.595*** | -0.334 |
| | (-2.201) | (2.216) | (-3.583) | (-1.767) |
| R-squared | 0.944 | 0.972 | 0.965 | 0.976 |

Note: Values included in parenthesis represent the robust standard errors; ***, **, and * represent 1%, 5%, and 10% significance levels, respectively.

(1) Comparative regression findings regarding the influence of low total GHG emissions groups' ESG performance on CFP.

(2) Regression results after introducing the interaction term between low total GHG emissions groups' ESG performance and KZ.

(3) Comparative regression findings regarding the influence of high total GHG emissions groups' ESG performance on CFP.

(4) Regression results after introducing the interaction term between high total GHG emissions groups' ESG performance and KZ.

## 6. Conclusion

Experts have varying viewpoints about the impact of ESG performance on CFP. Though prior researches have shown a positive relationship between ESG and CFP [21, 22], this study presents remarkable findings.

This research investigates the association between ESG and financial performance of the leading cruise companies worldwide from 2012 to 2023. It also investigates the moderate influence of financial constraints on this relationship, by a regression model. By addressing an area previously unexplored, this study improves the current research on the financial performance of the cruise sector. The key findings of the study are as follows:

ESG performance negatively impacts CFP in the cruise industry. However, this negative effect diminishes over time. Because of "black swan" events, the cruise companies have poor financial performance. Simultaneously, "green shipping" policy motivates them to invest more in reducing GHG emissions. Although these investments are costly and have a delayed impact on financial performance, causing temporary declines. the long-term investments in ESG projects may help cruise industry recover financially.

Furthermore, this research elaborates the correlation between ESG and CFP in different types of cruise companies. Through heterogeneity analysis, the results demonstrated that ESG has a more significant influence on CFP in cruise companies with large TIE ratios, as their sufficient financial resources and time support them to invest more in ESG. Moreover, cruise companies with low GHG emissions have a high correlation between ESG and CFP, because they have invested more to minimize emissions. This finding further validates the assumptions on the determinants of the negative relationship between ESG and CFP of cruise businesses.

This study also underlines that financial constraints of cruise companies have a detrimental moderating effect on ESG and CFP. Due to uneven access to information and funds, these constraints raise financing costs and discourage ESG initiatives, which redirects funds to operational needs and negatively affects ESG.

This report also offers advice for cruise operators and investors:

The cruise operators should explore diverse financing solutions to budget constraints, by combining internal and external resources. In terms of investing ESG projects, they should strategically allocate funds based on their commercial performance to avoid over-investment. To achieve operational efficiency, profitability, and environmental sustainability, they are expected to establish long-term strategic goals that may be achieved via a variety of activities, including scientific inquiry, technological progress, and energy conservation measures. By adopting this method, cruise operators can potentially encourage their economic value, meet public expectations for long-term development, and improve their status.

As for investors in the cruise industry, they should be aware that enterprises with strong ESG performance typically enjoy universal community support and mitigated investment risk. However, it is crucial to acknowledge that investing decisions should not be solely reliant on ESG ratings. Investors should assess CFP data from prior years and take into account the solvency, operational efficiency, profitability, and future growth opportunities. Investments should be tailored to individual needs and focus on long-term growth.

As the limitations exist in this paper due to the constraints of the available data, spaces remain for future research to fill in this gap. This study only draw conclusions about the correlation between the environmental aspects of cruise firms' ESG and CFP, based on the empirical facts related to GHG emissions and industry-wide emission reductions. Nevertheless, the evaluation of the environmental aspect includes metrics, while excluding the crucial metrics of social and governance in the study. This limitation may result in partial evaluation outcomes that do not accurately reflect the real cases. Hence, future research should emphasize on diverse data sources and conduct a more comprehensive analysis. Particularly, they should introduce precise evaluation system to thoroughly investigate the intricate connection between ESG and CFP in cruise industry, which will lay a more robust theoretical foundation for the sustainable development of the industry.